\documentclass[prl,twocolumn]{revtex4}

 \usepackage[english]{babel}
 \usepackage{epsfig,amssymb,amsmath}
 \usepackage{color}

\def\work{Letter\ }

\begin{document}

\title{Charge-memory polaron effect in molecular junctions}

\author{Dmitry\,A.~Ryndyk$^{1}$, Pino~D'Amico$^{1}$, Gianaurelio~Cuniberti$^{2}$, and Klaus~Richter$^{1}$}

\affiliation{$^{1}$Institute for Theoretical Physics, University of Regensburg,
D-93040 Regensburg, Germany \\ $^{2}$ Institute for Material Science and Max
Bergmann Center of Biomaterials, Dresden University of Technology, D-01062 Dresden,
Germany}

\begin{abstract}
The charge-memory effect, bistability and switching between charged and neutral
states of a molecular junction, as observed in recent STM experiments, is considered
within a minimal polaron model. We show that in the case of strong
electron-vibron interaction the rate of spontaneous quantum switching between
charged and neutral states is exponentially suppressed at zero bias voltage but can
be tuned through a wide range of finite switching timescales upon changing the bias.
We further find that, while junctions with symmetric voltage
drop give rise to random switching at finite bias, asymmetric junctions exhibit
hysteretic behavior enabling controlled switching. Lifetimes and charge-voltage curves
are calculated by the master equation method for weak coupling to the leads and at stronger
coupling by the equation-of-motion method for nonequilibrium Green functions.
\end{abstract}

\date{\today}
\maketitle

Memory effects and switching at the molecular scale are in the focus of present
experimental and theoretical studies within molecular electronics
\cite{Alexandrov03prb,Repp04science,Olsson07prl,Loertscher07prl,Liljeroth07science,DelValle07naturenanotech,Cuniberti05book}.
Beside stochastic switching in single-molecule junctions \cite{Loertscher07prl},
recent STM experiments \cite{Repp04science,Olsson07prl} show multistability of
neutral and charged states of single metallic atoms coupled to a metallic substrate
through a thin insulating ionic film. The switching was performed by the application
of a finite voltage to the STM tip and was explained by the large ionic
polarizability of the film \cite{Repp04science}.

The coupling of a charge to the displacement of ions in the film can be treated as
an electron-vibron interaction. If the energy of the {\em unoccupied} electron level
{\em without electron-vibron interaction} is $\epsilon_0$, the {\em occupied}
(charged) state of the {\em interacting system} will have the energy
$\epsilon_1=\epsilon_0-\epsilon_p$, where $\epsilon_p$ is so-called polaron shift
(or recombination energy). Neutral and charged (polaron) states correspond to local
minimums of the potential energy surface and are metastable, if the electron-vibron
interaction is strong enough. Applying an external voltage, one can change the state
of this bistable system, an effect that is accompanied by hysteretic charge-voltage
and current-voltage curves. In this approximation it is not necessary to include
Coulomb interaction explicitly, though one can additionally incorporate charging
effects.

\begin{figure}[b]
\begin{center}
\epsfxsize=0.65\hsize \epsfbox{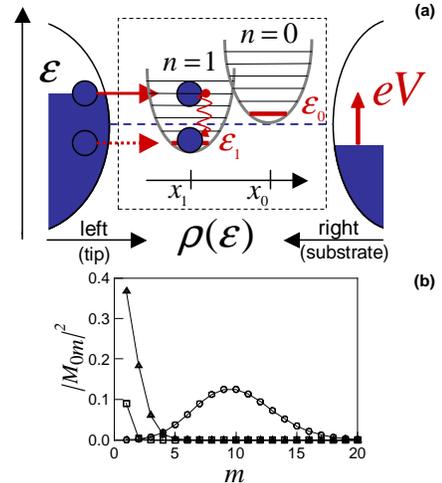}
\caption{(Color online) (a) The energy diagram of the
single-level electron-vibron model, coupled to left and right lead
(or tip and substrate in the case of STM). (b) Franck-Condon matrix elements $M_{0m}$
for weak ($g=0.1$, squares),
intermediate ($g=1$, triangles), and strong ($g=10$, circles) interaction.}
\label{V-SL-E}
\end{center}
\end{figure}

It was suggested \cite{Hewson79jphysc,Galperin05nanolett} that bistability between
charged and neutral states can be accounted for in a single-level model, when one
electron level is coupled to one vibration (Fig.\,\ref{V-SL-E}). The same problem
was also considered in Refs.~\cite{Mitra05prl,Mozyrsky06prb}, however with the
conclusion that quantum switching between bistable states results in telegraph
noise at finite voltage rather than in a memory effect. In this \work we show that
there is no contradiction among these two pictures, taking into account the
time-scale of the switching process. Indeed, the switching time $\tau$ between the
two states of interest should be compared with the characteristic time of the
external voltage sweeping, $\tau_s\sim V(t)/(dV(t)/dt)$. For $\tau\gg\tau_s$,
quantum switching can be neglected and hysteresis can be observed, while in the
opposite limit, $\tau\ll\tau_s$, the averaging removes the hysteresis. We calculate
the charge-voltage curves and describe the full crossover between two regimes.

The Hamiltonian of the single-level polaron model is
\begin{align}\label{H-SL}
& \hat H =(\epsilon_0+e\varphi_0)d^{\dag}d+\omega_0a^{\dag}a+
  \lambda\left(a^{\dag}+a\right)d^{\dag}d \nonumber \\
& +\sum_{ik}\left[(\epsilon_{ik}+e\varphi_i)
  c^{\dag}_{ik}c_{ik}+\left(V_{ik}c^{\dag}_{ik}d+h.c.\right)\right],
\end{align}
where the first three terms describe the free electron state, the free vibron of
frequency $\omega_0$ ($\hbar=1$) and the electron-vibron interaction. The further
terms are the Hamiltonian of the leads and the tunneling coupling ($i=L,R$ is the
lead index, $k$ labels electron states).
The electrical potential $\varphi_0$ plays an important role in
transport at finite bias voltages
$V=\varphi_L-\varphi_R$ between the left and right electrical potentials.
$\varphi_0$ describes the shift of the molecular level by the
bias voltage and can be written as $\varphi_0=\varphi_R+\eta(\varphi_L-\varphi_R)$,
$\eta\in[0,1]$ \cite{Datta97prl}.

The coupling to the leads is characterized by the level-width function
%
$
\Gamma_{i}(\epsilon)=2\pi\sum_{k}|V_{ik}|^2\delta(\epsilon-\epsilon_{ik}),
$
%
where the coupling $V_{ik}$ is assumed to be energy-independent (wide-band limit).
The full level broadening is given by the sum \mbox{$\Gamma=\Gamma_L+\Gamma_R$}.

Consider first the case of very weak coupling to the leads,
%
$
  \Gamma\ll\omega_0, \epsilon_p.
$
Using the polaron (Lang-Firsov)  \cite{Lang63jetp,Hewson74jjap,Mahan90book}
canonical transformation, it is easy to show that the eigenstates of the {\em
isolated system} ($\Gamma=0$) are
\begin{equation}\label{V-P-ES2}
  |\psi_{nm}\rangle=
  e^{-\frac{\lambda}{\omega_0}\left(a^\dag-a\right)d^\dag d}
  (d^\dag)^{n}\frac{(a^\dag)^{m}}{\sqrt{m!}}|0\rangle
\end{equation}
with the energies
\begin{equation}\label{V-P-EE}
  E_{nm}=\epsilon_1n+\omega_0m,\ \
  \epsilon_1=\epsilon_0-\frac{\lambda^2}{\omega_0},\ \ \epsilon_p=\frac{\lambda^2}{\omega_0}.
\end{equation}
When the system is weakly coupled to the leads, the polaron representation,
Eqs.~(\ref{V-P-ES2},\ref{V-P-EE}), is a convenient starting point. $n$ denotes the
number of electrons, while the quantum number $m$ characterizes vibronic
eigenstates, which are superpositions of states with different number of bare
vibrons. The qualitative picture of the sequential tunneling through a polaronic
state is given in Fig.\,\ref{V-SL-E}(a). Here the potential energies of the neutral
and charged states are sketched as a function of the vibronic coordinate $x$. When
the external voltage is applied, the energy levels are shifted depending on the
asymmetry parameter $\eta$. It should be noted that this type of the energy diagram
is quite general for charge-controlled bistable systems.

In the sequential tunneling regime the master equation for the probability
$p_{nm}(t)$ to find the system in one of the polaron eigenstates (\ref{V-P-ES2}) can
be written as \cite{Braig03prb,Mitra04prb,Koch05prl}
\begin{equation}\label{V-ST-ME}
  \frac{dp_{nm}}{dt}=\sum_{n'm'}\Gamma^{nn'}_{mm'}p_{n'm'}-
  \sum_{n'm'}\Gamma^{n'n}_{m'm}p_{nm}+I^V[p].
\end{equation}
Here the first term describes the tunneling transition {\em into the state} $|n,m
\rangle$ and the second term the transition {\em out of the state} $|n,m
\rangle$. $I^V[p]$ is the vibron scattering integral describing the relaxation of
vibrons to equilibrium. The transition rates $\Gamma^{nn'}_{mm'}$ are found from the
tunneling Hamiltonian (the last term in Eq.~(\ref{H-SL})).
Taking into account all possible single-electron tunneling processes, we obtain the
incoming and outgoing tunneling rates at zero bias voltage as
\begin{align}\label{Gin}
  \Gamma^{10}_{mm'} =\sum_{i=L,R}\Gamma_{i} & (E_{1m}-E_{0m'})\left|M_{mm'}\right|^2
  f_i^0(E_{1m}-E_{0m'}), \\
%
  \Gamma^{01}_{mm'} =
  \sum_{i=L,R}\Gamma_{i} & (E_{1m'}-E_{0m})\left|M_{mm'}\right|^2 \nonumber \\
  &\hskip 1cm \times\left(1-f_i^0(E_{1m'}-E_{0m})\right).
\end{align}
Here $f^0(\epsilon)$ is the equilibrium Fermi function, and
%
$
  M_{mm'}=\left\langle 0\left|\frac{a^{m}}{\sqrt{m!}}
  \exp\left[\frac{\lambda}{\omega_0}\left(a^\dag-a\right)\right]
  \frac{(a^\dag)^{m'}}{\sqrt{m'!}}\right|0\right\rangle
$
%
is the Franck-Condon matrix element. It is symmetric in $m-m'$ and can be calculated
analytically. For $m<m'$ it reads
\begin{equation}\label{V-ST-FC2}
  M_{m<m'}=\sum_{l=0}^{m}\frac{(-g)^l\sqrt{m!m'!}e^{-g/2}g^{(m'-m)/2}}
  {l!(m-l)!(l+m'-m)!},
\end{equation}
where $g=(\lambda/\omega_0)^2$ is the Huang-Rhys factor \cite{Huang50procrsoc}.

One characteristic feature of these matrix elements in transport is so-called
Franck-Condon blockade \cite{Koch05prl,Wegewijs05condmat}: in the case of strong
electron-vibron interaction the tunneling with small changes in $m$ is suppressed
exponentially, as illustrated in Fig.~\ref{V-SL-E}(b) for the matrix element
$M_{0m}=e^{-g/2}\frac{g^{m/2}}{\sqrt{m!}}$. Hence only tunneling through high-energy
states is possible. This is also suppressed at low bias voltage and low temperature.

Finally, the {\em average charge} is $\langle n\rangle(t)=\sum_{m}p_{1m}$, and the
{\em average current} (from the left or right lead) reads
$J_{i=L,R}(t)=e\sum_{mm'}\left(\Gamma^{10}_{imm'}p_{0m'}-\Gamma^{01}_{imm'}p_{1m'}\right)$.

\begin{figure}[t]
\begin{center}
\epsfxsize=0.75\hsize \epsfbox{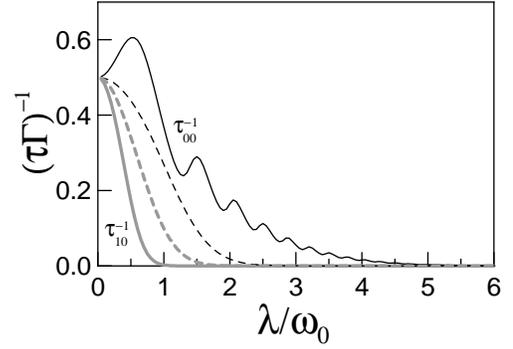}
\caption{Inverse life-time $(\tau\Gamma)^{-1}$ of the neutral state (thin solid line) and
the charged state (thick gray solid line)
as a function of $\lambda/\omega_0$ at $\epsilon_0=\lambda^2/2\omega_0$; and the same at
$\epsilon_0=0.9\lambda^2/\omega_0$ (dashed lines), $kT=0.1\omega_0$.}
\label{V-ST-M2}
\end{center}
\end{figure}

To proceed further, we calculate the characteristic life times of the neutral and
charged ground states. The life time $\tau_{nm}$ of the state $|n,m\rangle$ is
given by the sum of the rates of all possible processes which change this state,
%
$
  \tau^{-1}_{nm}=\sum_{n'm'}\Gamma^{n'n}_{m'm}.
$
%
As an example, calculating the life time of the neutral state $|0,0\rangle$, with an
energy higher than the charged ground state $|1,0\rangle$, we find
\begin{equation}\label{V-ST-T1}
  \tau^{-1}_{00}= \sum_{m}
  \sum_{i=L,R}\Gamma_{i}(E_{1m}-E_{00})\left|M_{m0}\right|^2
  f_i^0(E_{1m}-E_{00}).
\end{equation}
For energy-independent $\Gamma_i$ (the wide-band limit) we obtain the simple
analytical expression
\begin{equation}\label{V-ST-T2}
  \tau^{-1}_{00}=\Gamma\sum_{m}e^{-g}\frac{g^{m}}{m!}
  f^0\left(\epsilon_0-\frac{\lambda^2}{\omega_0}+\omega_0m\right).
\end{equation}
The corresponding expression for the life time of the charged state is (assuming
that the equilibrium electro-chemical potential in the leads is zero)
\begin{equation}\label{V-ST-T3}
  \tau^{-1}_{10}=\Gamma\sum_{m}e^{-g}\frac{g^{m}}{m!}
  f^0\left(-\epsilon_0+\frac{\lambda^2}{\omega_0}+\omega_0m\right).
\end{equation}

The dependence of the tunneling rates (\ref{V-ST-T2},\ref{V-ST-T3}) on the scaled
electron-vibron interaction constant $\lambda / \omega_0$ is shown in Fig.~\ref{V-ST-M2}. It is
clearly seen that at large values of $\lambda$ the tunneling from the neutral state
to the charged state and vice versa is exponentially suppressed in comparison with
the bare tunneling rate $\Gamma$. Hence both states are (meta)stable at low
temperatures and zero voltage.

Based on the experimental parameters of Ref.~\cite{Repp04science}, the charged
ground state is assumed to be below the equilibrium Fermi energy of the leads, while
the neutral ground state is above it. In the experiments \cite{Repp04science} the
observed relaxation energy $\epsilon_p\approx 2.4$ eV leads to the parameter
$\lambda/\omega_0$ of the order 5 to 10. Thus the system is in the blockade regime
at zero voltage, see Fig.~\ref{V-ST-M2}.

\begin{figure}[t]
\begin{center}
\epsfxsize=0.75\hsize \epsfbox{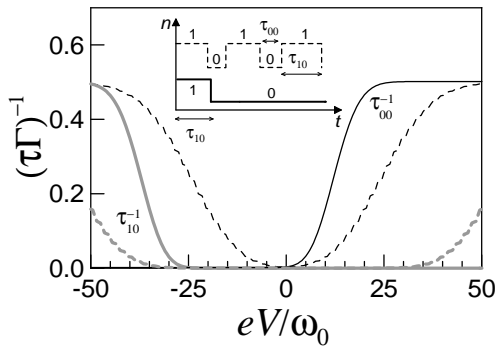}
\caption{Inverse life-time
$(\tau\Gamma)^{-1}$ as a function of normalized voltage $eV/\omega_0$ for the
asymmetric junction ($\eta=0$) at $\lambda/\omega_0=5$ and
$\epsilon_0=\lambda^2/2\omega_0$ for the neutral state (thin solid line), the
charged state (thick gray solid line) and the same for the symmetric junction
($\eta=0.5$, dashed lines). Inset: random switching between bistable states (dashed line)
and single switching into the stable state (full line) after a  sudden change of the voltage.}
\label{V-ST-M3}
\end{center}
\end{figure}

Next we consider the other important question, whether fast switching between the
two states is possible. At finite voltage the switching rates are
\begin{align}\nonumber
  \tau^{-1}_{00}=\sum_{m}\frac{e^{-g}g^{m}}{m!} & \left[
  \Gamma_Lf^0\left(\epsilon_1+\omega_0m-(1-\eta)eV\right)\right. \\ \label{V-ST-T4}
  & +\left.\Gamma_Rf^0\left(\epsilon_1+\omega_0m+\eta eV\right)
  \right], \\
%
\nonumber
  \tau^{-1}_{10}=\sum_{m}\frac{e^{-g}g^{m}}{m!} & \left[
  \Gamma_Lf^0\left(-\epsilon_1+\omega_0m+(1-\eta)eV\right)\right. \\ \label{V-ST-T5}
  & +\left.\Gamma_Rf^0\left(-\epsilon_1+\omega_0m-\eta eV\right)
  \right].
\end{align}

\begin{figure}[t]
\begin{center}
\epsfxsize=0.75\hsize \epsfbox{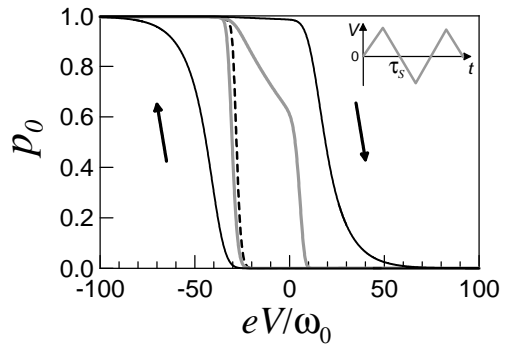}
\caption{Population of the neutral state
as a function of normalized voltage $eV/\omega_0$ in the asymmetrical junction
\mbox{($\eta=0$)} at $\lambda/\omega_0=5$ and
$\epsilon_0=\lambda^2/2\omega_0$ for fast voltage sweep (thin solid line), slower
sweep (thick gray solid line), and in the adiabatic limit (dashed line). Inset: sketch of
voltage time-dependence.}
\label{V-ST-M4}
\end{center}
\end{figure}

The voltage dependence of the inverse life time $(\tau\Gamma)^{-1}$ is shown in
Fig.~\ref{V-ST-M3} for a junction with the same tunneling
coupling, $\Gamma_L=\Gamma_R$, but asymmetric electrical field ($\eta=0$), as well
as for the completely symmetric junction ($\eta=0.5$).
The results in Fig.\ \ref{V-ST-M3} imply that in both cases
one can tune $(\tau\Gamma)^{-1}$ upon sweeping the bias voltage
and thereby control the timescales for switching between charged and neutral states.
For the symmetric junction both switching rates, $\tau_{00}^{-1}$ and
$\tau_{10}^{-1}$, (dashed lines) are simultaneously nonzero at finite voltage
($eV / \omega_0 \ge 40$ for the parameters of Fig.\ \ref{V-ST-M3}) leading to
random switching (noise) sketched as dashed line in the inset. On the contrary, for
the asymmetric junction controlled switching into the neutral (black solid line) and
charged (grey line) state can be achieved at large enough negative and positive voltage,
respectively. This qualitatively different
behaviour is a result of the distinct voltage asymmetry of the
two inverse lifetimes which are never both finite.
The further peculiar feature of the asymmetric case, namely that the
switching rates of the neutral and charged states interchange their role as
a function of bias, i.e., the neutral (charged) state is long-lived at negative
(positive) bias, implies hysteretic behavior and a memory effect.

To this end we consider what happens, if one sweeps the voltage with different velocity
(Fig.~\ref{V-ST-M4}) for the asymmetic case $\eta=0$. If the voltage is changed fast enough, i.e.\
faster than the
life time of charged and neutral states ($\tau\gg\tau_s$ as discussed in the
introduction), then both states can be obtained at zero voltage (hysteresis). In the
opposite (adiabatic) limit the change is so slow that the system relaxes into the
equilibrium state, and the population-voltage curve is single-valued. Note that this
controlled switching is possible only for asymmetric junctions for the reason given above.

\begin{figure}[t]
\begin{center}
\epsfxsize=0.75\hsize\epsfbox{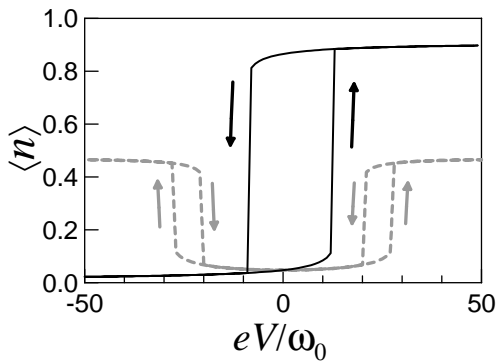}
\caption{Average number of electrons at $\Gamma_L=\Gamma_R=5\omega_0$
as a function of normalized voltage $eV/\omega_0$
for the asymmetric junction, $\eta=0$ (thin solid line), and for the
symmetric junction, $\eta=0.5$ (dashed line), for $\lambda/\omega_0=5$
and $\epsilon_0=\lambda^2/\omega_0$.}
\label{V-EOM1}
\end{center}
\end{figure}

We finally compare the results with those of a further important limiting case,
namely that the level width is finite (and possible finite dissipation of vibrons is
taken into account). Then the master equation approach can no longer be used, and we
apply alternatively the nonequilibrium Green function technique. Following Refs.\
\cite{Meir92prl,Haug96book}, the average number of electrons is determined by the
lesser Green function $G^<(t_1-t_2)=i\left\langle d^\dagger(t_2)d(t_1)\right\rangle$
as
%
$
\langle n\rangle=-i\int G^<(\epsilon)\frac{d\epsilon}{2\pi}.
$
%
The calculation of the Green function is a nontrivial task even in the single-level model.
It is simplified in the important limit of low vibron frequencies,
%
$
  \omega_0\ll\Gamma<\epsilon_p,
$
%
where the Born-Oppenheimer approximation holds true. We used the equation-of-motion
approach in this case.
In Fig.~\ref{V-EOM1} the charge-voltage dependence, obtained in the simplest
mean-field approximation \cite{Hewson79jphysc,Galperin05nanolett}, is shown. The
lesser function is represented as $G^<(\epsilon)=iA(\epsilon)f(\epsilon)$, with the
spectral and the distribution function
\begin{eqnarray}
  A(\epsilon) & = & \frac{2\Gamma}{(\epsilon-\epsilon_0-\frac{2\lambda^2}{\omega_0}\langle
  n\rangle-e\varphi_0)^2+\Gamma^2}, \\
f(\epsilon)& = & \frac{\Gamma_Lf_L^0(\epsilon-e\varphi_L)+\Gamma_Rf_R^0(\epsilon-e\varphi_R)}
{\Gamma_L+\Gamma_R} \, .
\end{eqnarray}
The result is qualitatively the same as in the sequential tunneling case: For
electrically asymmetric junctions two stable states exist at zero bias (memory effect),
which can be switched by the voltage.
The current shows similar hysteretic behaviour as a function of voltage.
For the symmetric junction hysteresis is observed only at finite voltage
(nonequilibrium bistability \cite{Galperin05nanolett}). Hence, asymmetric
junctions are again preferable for a memory effect.

Finally we note that in the case $\omega_0 \ll \Gamma$ we considered the stationary problem
only, assuming that the switching rate between the two metastable states is small
(compared e.g.\ to $\Gamma$) at large $\lambda / \omega_0$.
The calculation of the life times of metastable states
within the Green function appraoch and of dynamical effects arising
from the competition between voltage sweeping and switching times,
such as in Fig.~\ref{V-ST-M4}, remains as a problem for the future.

To conclude, we considered
a {\em charge-memory effect} and switching phenomena in single-molecule junctions
taking into account dynamical effects such as the interplay between timescales of
voltage sweeping and switching rates.
We showed that bistability arises if quantum transitions between neutral and
charged states involved are suppressed, e.g.\ due to Franck-Condon blockade.
Different regimes, characterized by random mutual transitions and by single switching
events into a stable configuration are identified. In the latter case controlled
switching of the molecule is achieved by applying finite voltage pulses.

We acknowledge fruitful discussions with J. Repp. This work was funded by the
Deutsche Forschungsgemeinschaft within the Priority Program SPP 1243 and
Collaborative Research Center SFB 689 (D.A.R.).


\begin{thebibliography}{24}
\expandafter\ifx\csname natexlab\endcsname\relax\def\natexlab#1{#1}\fi
\expandafter\ifx\csname bibnamefont\endcsname\relax
  \def\bibnamefont#1{#1}\fi
\expandafter\ifx\csname bibfnamefont\endcsname\relax
  \def\bibfnamefont#1{#1}\fi
\expandafter\ifx\csname citenamefont\endcsname\relax
  \def\citenamefont#1{#1}\fi
\expandafter\ifx\csname url\endcsname\relax
  \def\url#1{\texttt{#1}}\fi
\expandafter\ifx\csname urlprefix\endcsname\relax\def\urlprefix{URL }\fi
\providecommand{\bibinfo}[2]{#2}
\providecommand{\eprint}[2][]{\url{#2}}

\bibitem[{\citenamefont{Alexandrov and Bratkovsky}(2003)}]{Alexandrov03prb}
\bibinfo{author}{\bibfnamefont{A.~S.} \bibnamefont{Alexandrov}}
  \bibnamefont{and} \bibinfo{author}{\bibfnamefont{A.~M.}
  \bibnamefont{Bratkovsky}}, \bibinfo{journal}{Phys. Rev. B}
  \textbf{\bibinfo{volume}{67}}, \bibinfo{pages}{235312}
  (\bibinfo{year}{2003}).

\bibitem[{\citenamefont{Repp et~al.}(2004)\citenamefont{Repp, Meyer, Olsson,
  and Persson}}]{Repp04science}
\bibinfo{author}{\bibfnamefont{J.}~\bibnamefont{Repp}},
  \bibinfo{author}{\bibfnamefont{G.}~\bibnamefont{Meyer}},
  \bibinfo{author}{\bibfnamefont{F.~E.} \bibnamefont{Olsson}},
  \bibnamefont{and} \bibinfo{author}{\bibfnamefont{M.}~\bibnamefont{Persson}},
  \bibinfo{journal}{Science} \textbf{\bibinfo{volume}{305}},
  \bibinfo{pages}{493} (\bibinfo{year}{2004}).

\bibitem[{\citenamefont{Olsson et~al.}(2007)\citenamefont{Olsson, Paavilainen,
  Persson, Repp, and Meyer}}]{Olsson07prl}
\bibinfo{author}{\bibfnamefont{F.~E.} \bibnamefont{Olsson}},
  \bibinfo{author}{\bibfnamefont{S.}~\bibnamefont{Paavilainen}},
  \bibinfo{author}{\bibfnamefont{M.}~\bibnamefont{Persson}},
  \bibinfo{author}{\bibfnamefont{J.}~\bibnamefont{Repp}}, \bibnamefont{and}
  \bibinfo{author}{\bibfnamefont{G.}~\bibnamefont{Meyer}},
  \bibinfo{journal}{Phys. Rev. Lett.} \textbf{\bibinfo{volume}{98}},
  \bibinfo{eid}{176803} (\bibinfo{year}{2007}).

\bibitem[{\citenamefont{L{\"o}rtscher et~al.}(2007)\citenamefont{L{\"o}rtscher,
  Weber, and Riel}}]{Loertscher07prl}
\bibinfo{author}{\bibfnamefont{E.}~\bibnamefont{L{\"o}rtscher}},
  \bibinfo{author}{\bibfnamefont{H.~B.} \bibnamefont{Weber}}, \bibnamefont{and}
  \bibinfo{author}{\bibfnamefont{H.}~\bibnamefont{Riel}},
  \bibinfo{journal}{Phys. Rev. Lett.} \textbf{\bibinfo{volume}{98}},
  \bibinfo{pages}{176807} (\bibinfo{year}{2007}).

\bibitem[{\citenamefont{Liljeroth et~al.}(2007)\citenamefont{Liljeroth, Repp,
  and Meyer}}]{Liljeroth07science}
\bibinfo{author}{\bibfnamefont{P.}~\bibnamefont{Liljeroth}},
  \bibinfo{author}{\bibfnamefont{J.}~\bibnamefont{Repp}}, \bibnamefont{and}
  \bibinfo{author}{\bibfnamefont{G.}~\bibnamefont{Meyer}},
  \bibinfo{journal}{Science} \textbf{\bibinfo{volume}{317}},
  \bibinfo{pages}{1203} (\bibinfo{year}{2007}).

\bibitem[{\citenamefont{del Valle et~al.}(2007)\citenamefont{del Valle,
  Guti\'{e}rrez, Tejedor, and Cuniberti}}]{DelValle07naturenanotech}
\bibinfo{author}{\bibfnamefont{M.}~\bibnamefont{del Valle}},
  \bibinfo{author}{\bibfnamefont{R.}~\bibnamefont{Guti\'{e}rrez}},
  \bibinfo{author}{\bibfnamefont{C.}~\bibnamefont{Tejedor}}, \bibnamefont{and}
  \bibinfo{author}{\bibfnamefont{G.}~\bibnamefont{Cuniberti}},
  \bibinfo{journal}{Nature Nanotechnology} \textbf{\bibinfo{volume}{2}},
  \bibinfo{pages}{176} (\bibinfo{year}{2007}).

\bibitem[{\citenamefont{Cuniberti et~al.}(2005)\citenamefont{Cuniberti, Fagas,
  and Richter}}]{Cuniberti05book}
\bibinfo{author}{\bibfnamefont{G.}~\bibnamefont{Cuniberti}},
  \bibinfo{author}{\bibfnamefont{G.}~\bibnamefont{Fagas}}, \bibnamefont{and}
  \bibinfo{author}{\bibfnamefont{K.}~\bibnamefont{Richter}},
  \emph{\bibinfo{title}{Introducing Molecular Electronics}}
  (\bibinfo{publisher}{Springer-Verlag}, \bibinfo{year}{2005}).

\bibitem[{\citenamefont{Hewson and Newns}(1979)}]{Hewson79jphysc}
\bibinfo{author}{\bibfnamefont{A.~C.} \bibnamefont{Hewson}} \bibnamefont{and}
  \bibinfo{author}{\bibfnamefont{D.~M.} \bibnamefont{Newns}},
  \bibinfo{journal}{J. Phys. C: Solid State Phys.}
  \textbf{\bibinfo{volume}{12}}, \bibinfo{pages}{1665} (\bibinfo{year}{1979}).

\bibitem[{\citenamefont{Galperin et~al.}(2005)\citenamefont{Galperin, Ratner,
  and Nitzan}}]{Galperin05nanolett}
\bibinfo{author}{\bibfnamefont{M.}~\bibnamefont{Galperin}},
  \bibinfo{author}{\bibfnamefont{M.~A.} \bibnamefont{Ratner}},
  \bibnamefont{and} \bibinfo{author}{\bibfnamefont{A.}~\bibnamefont{Nitzan}},
  \bibinfo{journal}{Nano Lett.} \textbf{\bibinfo{volume}{5}},
  \bibinfo{pages}{125} (\bibinfo{year}{2005}).

\bibitem[{\citenamefont{Mitra et~al.}(2005)\citenamefont{Mitra, Aleiner, and
  Millis}}]{Mitra05prl}
\bibinfo{author}{\bibfnamefont{A.}~\bibnamefont{Mitra}},
  \bibinfo{author}{\bibfnamefont{I.}~\bibnamefont{Aleiner}}, \bibnamefont{and}
  \bibinfo{author}{\bibfnamefont{A.~J.} \bibnamefont{Millis}},
  \bibinfo{journal}{Phys. Rev. Lett.} \textbf{\bibinfo{volume}{94}},
  \bibinfo{pages}{076404} (\bibinfo{year}{2005}).

\bibitem[{\citenamefont{Mozyrsky et~al.}(2006)\citenamefont{Mozyrsky, Hastings,
  and Martin}}]{Mozyrsky06prb}
\bibinfo{author}{\bibfnamefont{D.}~\bibnamefont{Mozyrsky}},
  \bibinfo{author}{\bibfnamefont{M.~B.} \bibnamefont{Hastings}},
  \bibnamefont{and} \bibinfo{author}{\bibfnamefont{I.}~\bibnamefont{Martin}},
  \bibinfo{journal}{Phys. Rev. B} \textbf{\bibinfo{volume}{73}},
  \bibinfo{pages}{035104} (\bibinfo{year}{2006}).

\bibitem[{\citenamefont{Datta et~al.}(1997)\citenamefont{Datta, Tian, Hong,
  Reifenberger, Henderson, and Kubiak}}]{Datta97prl}
\bibinfo{author}{\bibfnamefont{S.}~\bibnamefont{Datta}},
  \bibinfo{author}{\bibfnamefont{W.}~\bibnamefont{Tian}},
  \bibinfo{author}{\bibfnamefont{S.}~\bibnamefont{Hong}},
  \bibinfo{author}{\bibfnamefont{R.}~\bibnamefont{Reifenberger}},
  \bibinfo{author}{\bibfnamefont{J.~I.} \bibnamefont{Henderson}},
  \bibnamefont{and} \bibinfo{author}{\bibfnamefont{C.~P.}
  \bibnamefont{Kubiak}}, \bibinfo{journal}{Phys. Rev. Lett.}
  \textbf{\bibinfo{volume}{79}}, \bibinfo{pages}{2530} (\bibinfo{year}{1997});
%
\bibinfo{author}{\bibfnamefont{T.}~\bibnamefont{Rakshit}},
  \bibinfo{author}{\bibfnamefont{G.-C.} \bibnamefont{Liang}},
  \bibinfo{author}{\bibfnamefont{A.~W.} \bibnamefont{Gosh}},
  \bibinfo{author}{\bibfnamefont{M.~C.} \bibnamefont{Hersam}},
  \bibnamefont{and} \bibinfo{author}{\bibfnamefont{S.}~\bibnamefont{Datta}},
  \bibinfo{journal}{Phys. Rev. B} \textbf{\bibinfo{volume}{72}},
  \bibinfo{pages}{125305} (\bibinfo{year}{2005}).

\bibitem[{\citenamefont{Lang and Firsov}(1963)}]{Lang63jetp}
\bibinfo{author}{\bibfnamefont{I.~G.} \bibnamefont{Lang}} \bibnamefont{and}
  \bibinfo{author}{\bibfnamefont{Y.~A.} \bibnamefont{Firsov}},
  \bibinfo{journal}{Sov. Phys. JETP} \textbf{\bibinfo{volume}{16}},
  \bibinfo{pages}{1301} (\bibinfo{year}{1963}).

\bibitem[{\citenamefont{Hewson and Newns}(1974)}]{Hewson74jjap}
\bibinfo{author}{\bibfnamefont{A.~C.} \bibnamefont{Hewson}} \bibnamefont{and}
  \bibinfo{author}{\bibfnamefont{D.~M.} \bibnamefont{Newns}},
  \bibinfo{journal}{Japan. J. Appl. Phys.} \textbf{\bibinfo{volume}{Suppl.\,2,
  Pt.\,2}}, \bibinfo{pages}{121} (\bibinfo{year}{1974}).

\bibitem[{\citenamefont{Mahan}(1990)}]{Mahan90book}
\bibinfo{author}{\bibfnamefont{G.}~\bibnamefont{Mahan}},
  \emph{\bibinfo{title}{Many-Particle Physics}} (\bibinfo{publisher}{Plenum,
  N. Y.}, \bibinfo{year}{1990}).

\bibitem[{\citenamefont{Braig and Flensberg}(2003)}]{Braig03prb}
\bibinfo{author}{\bibfnamefont{S.}~\bibnamefont{Braig}} \bibnamefont{and}
  \bibinfo{author}{\bibfnamefont{K.}~\bibnamefont{Flensberg}},
  \bibinfo{journal}{Phys. Rev. B} \textbf{\bibinfo{volume}{68}},
  \bibinfo{pages}{205324} (\bibinfo{year}{2003}).

\bibitem[{\citenamefont{Mitra et~al.}(2004)\citenamefont{Mitra, Aleiner, and
  Millis}}]{Mitra04prb}
\bibinfo{author}{\bibfnamefont{A.}~\bibnamefont{Mitra}},
  \bibinfo{author}{\bibfnamefont{I.}~\bibnamefont{Aleiner}}, \bibnamefont{and}
  \bibinfo{author}{\bibfnamefont{A.~J.} \bibnamefont{Millis}},
  \bibinfo{journal}{Phys. Rev. B} \textbf{\bibinfo{volume}{69}},
  \bibinfo{pages}{245302} (\bibinfo{year}{2004}).

\bibitem[{\citenamefont{Koch and von Oppen}(2005)}]{Koch05prl}
\bibinfo{author}{\bibfnamefont{J.}~\bibnamefont{Koch}} \bibnamefont{and}
  \bibinfo{author}{\bibfnamefont{F.}~\bibnamefont{von Oppen}},
  \bibinfo{journal}{Phys. Rev. Lett.} \textbf{\bibinfo{volume}{94}},
  \bibinfo{pages}{206804} (\bibinfo{year}{2005});
%
\bibinfo{author}{\bibfnamefont{J.}~\bibnamefont{Koch}},
  \bibinfo{author}{\bibfnamefont{M.}~\bibnamefont{Semmelhack}},
  \bibinfo{author}{\bibfnamefont{F.}~\bibnamefont{von Oppen}},
  \bibnamefont{and} \bibinfo{author}{\bibfnamefont{A.}~\bibnamefont{Nitzan}},
  \bibinfo{journal}{Phys. Rev. B} \textbf{\bibinfo{volume}{73}},
  \bibinfo{pages}{155306} (\bibinfo{year}{2006}{\natexlab{a}}).

\bibitem[{\citenamefont{Nowack and Wegewijs}(2003)}]{Wegewijs05condmat}
\bibinfo{author}{\bibfnamefont{K.~C.}~\bibnamefont{Nowack}} \bibnamefont{and}
  \bibinfo{author}{\bibfnamefont{M.~R.}~\bibnamefont{Wegewijs}},
  \bibinfo{journal}{cond-mat/0506552}.

\bibitem[{\citenamefont{Huang and Rhys}(1950)}]{Huang50procrsoc}
\bibinfo{author}{\bibfnamefont{K.}~\bibnamefont{Huang}} \bibnamefont{and}
  \bibinfo{author}{\bibfnamefont{A.}~\bibnamefont{Rhys}},
  \bibinfo{journal}{Proc. R. Soc. London Ser. A}
  \textbf{\bibinfo{volume}{204}}, \bibinfo{pages}{406} (\bibinfo{year}{1950}).

\bibitem[{\citenamefont{Meir and Wingreen}(1992)}]{Meir92prl}
\bibinfo{author}{\bibfnamefont{Y.}~\bibnamefont{Meir}} \bibnamefont{and}
  \bibinfo{author}{\bibfnamefont{N.~S.} \bibnamefont{Wingreen}},
  \bibinfo{journal}{Phys. Rev. Lett.} \textbf{\bibinfo{volume}{68}},
  \bibinfo{pages}{2512} (\bibinfo{year}{1992});
%
\bibinfo{author}{\bibfnamefont{A.-P.} \bibnamefont{Jauho}},
  \bibinfo{author}{\bibfnamefont{N.~S.} \bibnamefont{Wingreen}},
  \bibnamefont{and} \bibinfo{author}{\bibfnamefont{Y.}~\bibnamefont{Meir}},
  \bibinfo{journal}{Phys. Rev. B} \textbf{\bibinfo{volume}{50}},
  \bibinfo{pages}{5528} (\bibinfo{year}{1994}).

\bibitem[{\citenamefont{Haug and Jauho}(1996)}]{Haug96book}
\bibinfo{author}{\bibfnamefont{H.}~\bibnamefont{Haug}} \bibnamefont{and}
  \bibinfo{author}{\bibfnamefont{A.-P.} \bibnamefont{Jauho}},
  \emph{\bibinfo{title}{Quantum Kinetics and Optics of Semiconductors}}, vol.
  \bibinfo{volume}{123} of \emph{\bibinfo{series}{Springer Series in
  Solid-State Sciences}} (\bibinfo{publisher}{Springer}, \bibinfo{year}{1996}).

\end{thebibliography}
\end{document}